# A Unified AI System for Data Quality Control and DataOps Management in Regulated Environments

**Devender Saini**
BlackRock, Inc.
New York, NY, USA

**Bhavika Jain**
BlackRock, Inc.
New York, NY, USA

**Nitish Ujjwal**
BlackRock, Inc.
New York, NY, USA

**Philip Sommer**
BlackRock, Inc.
New York, NY, USA

**Dan Romuald Mbanga**
BlackRock, Inc.
New York, NY, USA

**Dhagash Mehta**
BlackRock, Inc.
New York, NY, USA

## Abstract

In regulated domains such as finance, the integrity and governance of data pipelines are critical—yet existing systems treat data quality control (QC) as an isolated preprocessing step rather than a first-class system component. We present a unified AI-driven Data QC and DataOps Management framework that embeds rule-based, statistical, and AI-based QC methods into a continuous, governed layer spanning ingestion, model pipelines, and downstream applications. Our architecture integrates open-source tools with custom modules for profiling, audit logging, breach handling, configuration-driven policies, and dynamic remediation. We demonstrate deployment in a production-grade financial setup: handling streaming and tabular data across multiple asset classes and transaction streams, with configurable thresholds, cloud-native storage interfaces, and automated alerts. We show empirical gains in anomaly detection recall, reduction of manual remediation effort, and improved auditability and traceability in high-throughput data workflows. By treating QC as a system concern rather than an afterthought, our framework provides a foundation for trustworthy, scalable, and compliant AI pipelines in regulated environments.

## 1 Introduction

In regulated industries, data quality and operational reliability form the basis for compliance, transparency, and accountability. Sectors such as healthcare, energy, and finance rely on complex data pipelines that are subject to strict oversight, where even minor inconsistencies can lead to compliance breaches, financial losses, or reputational risks. Among these, the financial industry faces some of the most demanding challenges, as institutions must process large volumes of heterogeneous data across markets, products, and counterparties while adhering to evolving regulatory frameworks. In this context, the integration of DataOps (Ereth, 2018) with rigorous data quality management has become essential. Financial institutions increasingly depend on high-frequency, multi-source data to inform investment decisions, assess risk, enhance client experience, and maintain continuous regulatory compliance. In the remainder of this paper, though we will focus mostly on the financial industry, the discussion and the proposed framework are directly applicable to any other regulated industry such as insurance, healthcare, defense, government, etc.

To design a robust DataOps and quality control framework, it is first necessary to clarify what constitutes "data" and data quality issues in regulated environments. Understanding the diverse forms, sources, and transformations of data provides the foundation for identifying quality challenges and establishing appropriate validation and monitoring mechanisms.

### 1.1 What is data, and what is not data?

*Data* refers to any collection of values or facts that can be processed and analyzed to support decision-making (Fan & Geerts, 2022). It encompasses numerical, textual, and categorical information generated through observations, measurements, or transactions. Financial institutions handle vast and diverse data, including financial statements (balance sheets, income statements, cash flow reports), market data (historical and real-time prices, indices, and transactions), client and transaction records (personal details, credit ratings, trading histories), technical documentation (blueprints, software specifications), and risk and compliance data (metadata, regulatory reports, data lineage). These datasets underpin activities such as trading, portfolio management, risk modeling, regulatory compliance, and governance.

Most financial data are structured in tabular form, organized in rows and columns within databases, data warehouses, or spreadsheets. Streaming tabular data, by contrast, are generated and processed continuously in real time, such as stock feeds, transaction logs, or financial news updates. Data sources include stock exchanges, central banks, market data vendors, public repositories, and internal enterprise systems, as well as outputs from predictive and analytical models. This paper focuses on tabular and streaming tabular data, while extending the framework to other structured and unstructured modalities is left for future work.

A *model* is a quantitative method that processes input data into quantitative estimates using statistical, economic, financial, or mathematical theories (Hasan et al., 2020; FederalReserveSystem, 2011). Although models are constructed from data, their parameters and structure can be considered metadata or hyperparameters. The model can be thought of as a data-processor that takes a set of input data and transforms it to another dataset. This gives birth to a conceptual question about the true origin and validation scope of data. In financial systems, the output of one model frequently serves as the input to another. Hence, Data QC must operate at multiple boundaries, both on the raw inputs and on the outputs of intermediate models. When a model's output serves as input data for another model, the QC process may be called as *model surveillance* from the organizational perspective, though a well thought-through Data QC and DataOps Management framework must play the dual role of performing Data QC as well as model surveillance.

### 1.2 Types and Applications of Data Quality Control in the Financial Systems

Ensuring high data quality is a prerequisite for developing reliable, transparent, and explainable AI systems in regulated domains. Within the financial industry, this requirement becomes even more critical due to the heterogeneous, high-velocity, and interdependent nature of financial data. Information used for trading, valuation, risk management, and regulatory reporting often originates from multiple internal and external sources, each with distinct structures, update frequencies, and levels of reliability. Any degradation in data accuracy, completeness, or timeliness can propagate through automated and analytical models, producing biased insights, unstable model behavior, and costly operational errors. As previous studies have shown, inadequate data quality not only undermines model performance but can also result in flawed risk assessments and substantial financial losses (Mahanti, 2019; Liu, 2020).

Table 3 summarizes these key categories of data quality issues, their typical manifestations, and their potential consequences.

Below we outline key applications of a rigorous Data QC framework within financial systems:

- **Risk Management:** Data QC plays a critical role in financial risk management by identifying incorrect or corrupted data points before they are used in decision-making or modeling processes (Tatineni & Mustyala, 2024). A less recognized source of model risk is the extrapolation behavior of AI algorithms, which are universal function approximators (Cybenko, 1989; Hornik, 1991; Augustine, 2024) but are only guaranteed to approximate the ground truth within a bounded data region. When new data fall outside this range, predictions become unreliable (Cao & Yousefzadeh, 2023). Simple rule-based QC checks, such as flagging data points that deviate from the prior day's mean by more than a defined threshold, can help prevent such extrapolation errors.

- **Signal Generation and Fraud Detection:** Data flagged as anomalous by QC mechanisms, even if not erroneous or corrupted, may indicate potential fraud (e.g., suspicious credit card transactions) or emerging market signals (Ahmed et al., 2016; Hilal et al., 2022). This dual interpretation makes anomaly detection an integral component of both operational and strategic decision-making.

- **Model Diagnostics and Enhancement:** A Data QC framework can identify data subsets where model performance deteriorates, allowing modelers to examine feature distributions and detect sources of bias. This supports targeted feature engineering and model refinement, improving accuracy and fairness (Zhang et al., 2018).

- **Model Surveillance:** Continuous monitoring of model performance metrics and key indicators helps detect performance drift or structural anomalies. When embedded within the Data QC pipeline, this surveillance function ensures long-term model robustness and regulatory compliance.

### 1.3 Stages of Data Quality Control

Data QC must operate as a continuous process throughout the lifecycle of a financial model. It begins with validating input data received from internal or third-party sources, extends to verifying the correctness of model outputs, and concludes with post-processing checks that ensure reliability before the data is consumed by downstream systems. As illustrated in Figure 1, these stages are interconnected and form the backbone of a trustworthy DataOps and AIOps ecosystem. Automating these steps is essential because data quality, model performance, and system governance are tightly coupled.

We envision a rigorous data QC process organized into a two-tier process consisting of Centralized QC and Model-Level (Context-Aware) QC, which together cover both upstream and downstream quality assurance.

1. Centralized QC: This layer operates at the enterprise or data platform level, immediately after data ingestion from external or internal sources. Its objective is to enforce consistency, standardization, and compliance across organizational boundaries. Centralized QC focuses on upstream processes such as schema validation, format verification, deduplication, reference data reconciliation, and regulatory reporting checks. For example, before a dataset from a market vendor is made available to analysts, centralized QC ensures that all files comploy with the enterprise schema, timestamps are in the expected format, and all mandatory fields (such as CUSIP or ISIN identifiers) are present. These checks are typically integrated within Exploration, Transform and Load (ETL) workflows or data lake pipelines, ensuring that only high-quality curated data flows into modeling environments.

2. Model-Level (Context-Aware) QC: Once data reaches a specific modeling or analytical context, additional quality checks are required that reflect the model's purpose and sensitivity. This stage applies QC within individual inference pipelines, focusing on model-specific inputs, intermediate outputs, and post-processing results. Model-level QC detects local anomalies that may bypass enterprise validation, such as feature drift, input–output inconsistencies, or domain-specific constraint violations. For instance, an asset pricing model may flag a breach if the mean spread between corporate and government bond yields deviates beyond a statistically defined tolerance, even if the data passed centralized QC. Similarly, a credit risk model may identify an abnormal concentration of missing values in a key explanatory variable, triggering a model-specific imputation policy. This form of QC is therefore context-aware and directly tied to model governance and performance monitoring.

Centralized QC provides a standardized, organization-wide foundation for data integrity, while model-level QC delivers adaptive, fine-grained controls that align with specific analytical objectives. Both layers complement each other: the first ensures uniformity and regulatory compliance across datasets, and the second enables dynamic surveillance of model health under evolving market conditions. Embedding these QC layers within the model lifecycle allows practitioners to distinguish between performance degradation caused by data issues and those arising from shifts in market dynamics, thereby supporting timely and targeted interventions.

To summarize:

1. Upstream QC validates raw and curated data entering the modeling pipeline, addressing issues such as missing fields, duplicate records, or incorrect data types.

2. Model QC performs real-time, context-specific quality checks within individual inference workflows to detect data drift, anomalies, or structural inconsistencies.

3. Downstream QC evaluates the reliability and coherence of model outputs before they are propagated to business systems or subsequent analytical layers, ensuring traceability and consistency across the data lifecycle.

Figures 6 and 7 in the Appendix illustrate the main steps involved in centralized and model-level data quality control processes.

The multi-layer data QC paradigm reflects the practical reality that even when upstream teams enforce enterprise-wide standards, local anomalies, drifts, or missing values can still emerge at the model stage. By embedding quality checks directly into the model life-cycle, we create a safeguard that can distinguish between performance issues caused by data problems versus those caused by market condition changes, enabling more precise interventions.

## 2 LITERATURE REVIEW

Data QC is a foundational requirement for reliable AI and data-driven decision-making, yet an end-to-end, continuously governed framework that combines traditional and modern QC approaches remains limited in the literature (Sambasivan et al., 2021). Existing research defines data quality along dimensions such as accuracy, completeness, consistency, and timeliness (Mahanti, 2019; Jain et al., 2020), but these dimensions are usually treated as static governance concepts rather than operational processes embedded within the DataOps lifecycle.

Recent works have explored integrated data pipelines as engines for model deployment and decision support. Karim (Ramin et al., 2023) introduced the concept of an *AI factory*, encompassing data pipelines, algorithm development, and experimentation platforms as interconnected components. Iansiti and Lakhani (Iansiti & Lakhani, 2020) further emphasized that standardized and automated pipelines improve business agility and scalability. However, these efforts primarily focus on workflow efficiency rather than on continuous data validation and governance.

Traditional QC frameworks rely on rule-based validation or statistical checks implemented as isolated preprocessing steps. Zha et al. (Zha et al., 2018) surveyed quality improve-ment methods in AI workflows, emphasizing schema validation and integrity constraints. Several open-source tools have emerged to address these needs. TensorFlow Data Validation (TFDV) (TensorFlow, 2024) provides large-scale schema enforcement and anomaly detection capabilities within the TensorFlow Extended (TFX) ecosystem (Caveness & GC, 2020). Great Expectations (GEx) (Great Expectations, 2024) offers a declarative syntax for writing human-readable "expectations" and integrates easily with CI/CD pipelines, allowing reproducible, test-driven data validation.

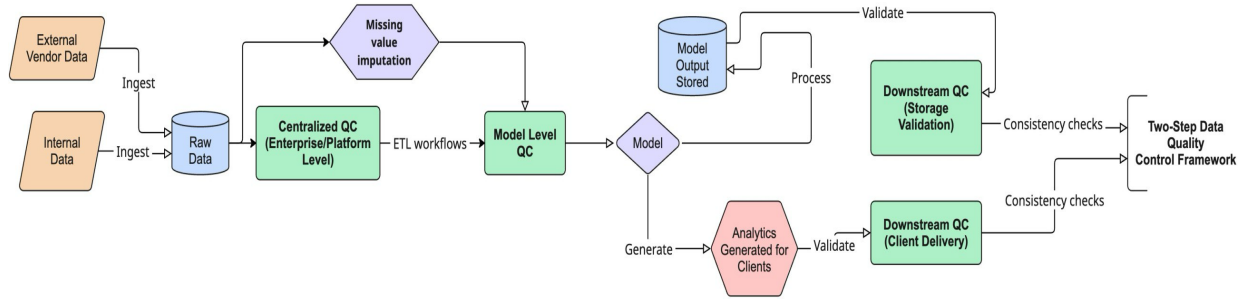

*Figure 1.* Two-Step Data QC Framework is enforced at multiple layers: Centralized QC ensures enterprise-level data hygiene, Model QC safeguards inference pipelines, and Downstream QC validates results before consumption.

Beyond rule-based methods, research has increasingly focused on automating QC through AI algorithms for anomaly detection and imputation. Dreves et al. (Dreves et al., 2021) and Whang et al. (Whang et al., 2023) demonstrated that AI models can proactively identify data discrepancies before they propagate downstream. Yandrapalli et al. (Yandrapalli, 2024) compared statistical and AI-based techniques such as Tukey's Interquartile Range, Isolation Forest, and DBSCAN, highlighting trade-offs between interpretability and adaptability. Other relevant studies include Nonnenmacher et al. (Nonnenmacher & Gómez, 2021), who examined automated anomaly detection pipelines; Nassif et al. (Nassif et al., 2021), who categorized anomaly detection methods by learning type; and Bakumenko et al. (Bakumenko & Elragal, 2022), who applied AI-based QC to fraud detection in financial services.

Collectively, these works underscore a growing shift toward automation and continuous monitoring in data quality assurance. However, none integrate rule-based, statistical, and AI-based QC within a single governed architecture. The present study addresses this gap by proposing a unified DataOps framework that combines these paradigms into a transparent, auditable, and scalable system designed for regulated environments such as financial services.

### 2.1 Our Contributions

This paper presents a unified and continuously governed approach to Data QC within the DataOps life-cycle. The main contributions are:

- **Unified Framework:** We design and implement an end-to-end Data QC and DataOps Management Framework that integrates rule-based validation, AI-based anomaly detection, and imputation-aware checks into one cohesive system.

- **Tool Integration:** The framework combines TensorFlow Data Validation (TensorFlow, 2024), Great Expectations (Great Expectations, 2024), and PyOD (Zhao et al., 2019), enhanced with custom modules for profiling, audit logging, breach handling, and stakeholder notifications. The choice of open-source libraries is deliberate in order to ensure transparency, reproducibility, and ease of integration within existing enterprise DataOps ecosystems.

- **Continuous and Governed QC:** QC is embedded as a continuous, governed layer within the DataOps life-cycle, enabling proactive monitoring and remediation.

- **Compliance and Scalability:** Implemented as a scalable, platform-agnostic Python library following National Institute of Standards and Technology (NIST) guidelines (NIST., 2008), the system supports modular deployment across regulated domains.

- **Operational Impact:** The system provides QC-driven observability across data and models, ensuring reliable and auditable validation in financial AI workflows.

In the next Section, we describe the proposed framework and provide details of our implementation of the system.

# 3 PROPOSED DESIGN AND IMPLEMENTATION DETAILS

The system consists of two primary components: a core backend library that implements rule-based, statistical, and AI-driven data quality functions, and a front-end layer that integrates these capabilities into production models and workflows. In a typical production setup, the process is often orchestrated through scheduled automation and includes data ingestion, feature generation, and immediate application of QC checks before or after transformation, depending on the use case. Subsequent steps, such as computing the derived variables or similarity scores, are followed by targeted QC validation. The framework supports seamless integration with cloud platforms, enabling efficient deployment and maintenance. Thresholds for breach categories can be dynamically adjusted without redeployment, ensuring flexibility and production-grade reliability.

## 3.1 Back-End Process

The backend is implemented as a lightweight Python library centered on a QC execution engine that orchestrates the entire quality control process. This core component manages both current and historical data paths, applies QC rules to the input datasets as outlined in Table 4 (see Appendix), and records breach results in a designated repository. It also triggers automated notifications when data is missing or when threshold violations are detected.

Above this core engine lies a data validation layer responsible for statistical and schema-based validation using a comparison module built on TFDV and GEx. Alternatively, users can specify a customized configuration file containing additional profiling functions that can be executed at the front end. Figure 2 illustrates the workflow of the proposed framework, showing how it performs both standard and advanced rule-based as well as statistical QC checks.

The backend also includes a set of generic helper utilities that manage file input and output across multiple formats (CSV, JSON, Parquet), handle path resolution, and perform data saving, logging, and performance tracking. Through the Storage Interface Layer, the framework supports seamless interaction with local file systems and major cloud storage platforms such as AWS Simple Storage Service (S3)[1] and Google Cloud Storage (GCS)[2], ensuring flexible and scalable data management.

## 3.2 Front-End Process

The front end provides the interface through which QC functions are executed. These functions can be imported directly from the custom Python library or invoked from a configuration file. Each QC model includes a central script that serves as the main entry point. This script loads configuration parameters, identifies the relevant QC specification, and calls the corresponding logic based on model type or input arguments.

The specification layer contains multiple spec modules, each designed for a particular data domain such as equities, corporate bonds, mutual funds, etc. These modules define domain-specific QC checks and include logic for handling features unique to that domain, such as identifiers or scoring metrics. Shared utility functions are used to perform standard checks including null detection, duplicate identification, threshold validation, and schema comparison. A Spec Dispatcher dynamically maps specification names to their corresponding classes, allowing the main execution file to remain generic and scalable across diverse QC workflows. The core functionality layer provides reusable components that support all spec modules through a superclass responsible for common tasks such as path resolution, breach detection, and result storage.

Configuration management is handled through a resource file that stores runtime parameters such as input and output paths, threshold values, and email notification settings. These parameters are loaded dynamically, allowing flexible customization across different environments without redeployment. Once the QC checks are completed, breach results are compiled automatically and alerts are generated. The system produces summary reports, breach notification emails, and dashboard feeds that consume flat-file outputs to ensure consistent data quality across the pipeline. Breaches are categorized by severity and type, which allows for differentiated responses. Severe breaches halt the process and trigger immediate alerts, while less critical issues only generate notifications. Breach data can also be stored for visualization in monitoring dashboards.

A comprehensive HTML report is generated at the end of each run, summarizing QC statistics and describing the data distribution. Each feature in the dataset is profiled to provide summary statistics. Numeric features include metrics such as count, mean, median, and standard deviation, whereas categorical features report unique values, frequencies, and average string lengths.

The framework supports three primary types of QC: rule-based, statistical, and AI-based. In addition, an imputation-aware, model-based QC component is included to enhance completeness and consistency across the pipeline.

### 3.2.1 Rule-Based Validation

The rule-based data quality control (QC) methodology applies predefined validation rules to ensure systematic and repeatable assessment of data integrity. It supports multiple data formats and leverages the TensorFlow Data Validation (TFDV) library (TensorFlow, 2024) to detect anomalies and outliers by comparing dataset statistics against a defined schema. The system also evaluates data drift and skewness, helping maintain the reliability and stability of downstream models. Customized reports can be generated for any detected breaches, allowing users to trace and address quality issues efficiently.

---

[1] https://aws.amazon.com/s3/
[2] https://cloud.google.com/storage

To extend the capabilities of the QC framework, GEx is incorporated as a complementary validation engine alongside TensorFlow Data Validation. GEx provides a flexible domain-specific language for defining human-readable, version-controlled data assertions such as `expect_column_values_to_be_between`, `expect_table_row_count_to_be_between`, and `expect_column_values_to_not_be_null`. It also generates rich "Data Docs" for reporting validation results. GEx operates across multiple execution backends including Pandas, Spark, and SQL, which makes it adaptable to heterogeneous enterprise environments. Its checkpoint and action-based model, combined with CI/CD-friendly integrations, enables automated DataOps gates that promote continuous quality assurance.

Note that the rule-based data QC can be even more crucial for the centralized QC done at the enterprise level. Table 4 summarizes the complete set of rule-based data quality checks implemented in our system.

### 3.2.2 *Statistical Validation*

The second methodology in the framework is statistical-based QC, which applies statistical techniques to evaluate whether data points fall within acceptable limits (Ryan, 2011). This approach focuses on identifying outliers and distributional irregularities that may indicate data quality issues. Several statistical methods can be employed for outlier detection. The *max–min approach* defines a valid range using the maximum and minimum values of a dataset, flagging any observations outside this range as potential outliers. The *percentile approach* uses thresholds such as the 95th or 99th percentile to capture extreme values irrespective of the overall distribution shape (Aggarwal, 2017). The *modified Z-score method* is suitable for datasets that do not follow a normal distribution, as it adjusts for skewness and scale effects (Venkataanusha et al., 2019). Finally, *Tukey's fences* define boundaries around the interquartile range (IQR), identifying points outside these limits as possible outliers (Iftikhar & Ateeq, 2015). These methods enable comprehensive detection of statistical anomalies that could affect model performance or downstream analyses.

### 3.3 AI-Based Advanced QC

### 3.3.1 *AI-Based Quality Control*

AI-based QC employs learning algorithms to identify anomalies, detect missing or inconsistent patterns, and improve overall data integrity (Bakumenko & Elragal, 2022). Unlike rule-based or statistical QC, which rely on predefined thresholds or distributional assumptions, AI-driven approaches can learn complex, non-linear relationships directly from data. This enables the system to capture context-dependent irregularities that traditional methods often overlook, particularly in dynamic and high-dimensional datasets. As a result, AI-based QC provides greater adaptability, scalability, and precision in maintaining data reliability across continuously evolving data pipelines.

AI algorithms commonly fall into four categories: supervised, unsupervised, semi-supervised, and reinforcement learning (Ayodele, 2010). These methods can be used for both anomaly detection and intelligent imputation of missing values. Effective management of missing data is essential for ensuring robust analytics, with methods such as predictive imputation or tree-based models like Random Forest offering data-driven solutions (Horton & Kleinman, 2007; Emmanuel et al., 2021).

In this framework, the PyOD library (Zhao et al., 2019) is used as the primary engine for AI-based QC. PyOD is an open-source Python toolkit that provides more than fifty anomaly detection algorithms, supporting both batch and online detection modes. It offers scalability, extensibility, and seamless integration with existing AI workflows. The library supports multiple data modalities including time series, graph, and tabular data, making it suitable for complex financial datasets. Selecting the appropriate PyOD model depends on the data type and the intended analytical objective, ensuring that anomaly detection remains both accurate and computationally efficient. Table 5 summarizes the set of AI algorithms integrated into the system and their corresponding applications for data quality assurance.

Before training the anomaly detection model, variables unsuitable for modeling are transformed through normalization, encoding of categorical attributes, or feature scaling and anomalies are corrected based on the results captured from the input data QC stage. The prepared dataset is then randomly divided into training and testing subsets, with the training portion further split into training and validation sets. Hyperparameter tuning is conducted on the validation set to optimize model performance and prevent overfitting.

During model training, each data instance is assigned an anomaly score that quantifies its deviation from the learned normal patterns. A threshold is then calibrated to differentiate between normal and anomalous observations. After training, the model is evaluated on the test dataset using appropriate performance metrics depending on the chosen AI algorithm. Once anomalous records are identified, they are flagged for review and remediation based on the operational data cadence, which may be daily, weekly, or monthly. This process, illustrated in Figure 3, ensures that only high-quality and trustworthy data progresses to downstream applications.

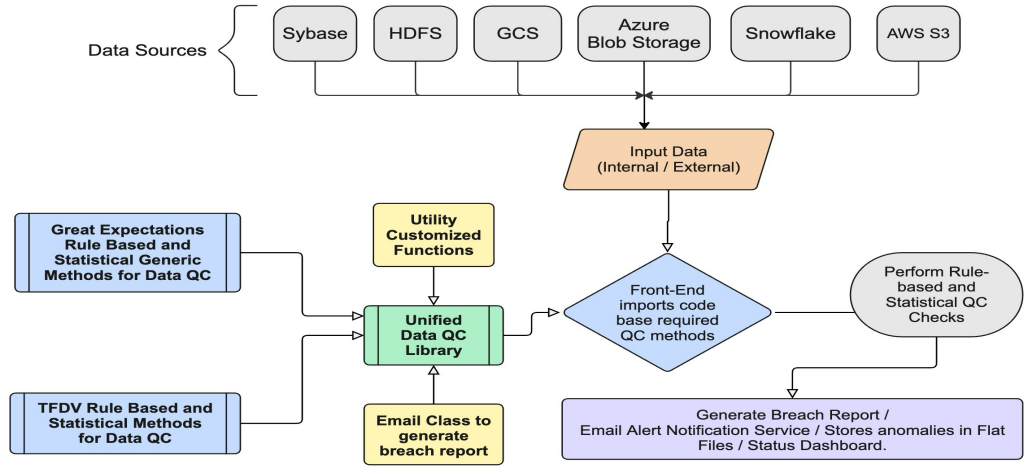

*Figure 2.* Rule-Based and Statistical Data Quality Assessment Application Flow Chart

### 3.4 Imputation-Aware Model-Based QC

Handling missing data is not merely a pre-processing task but a central component of the system. It is performed immediately after raw data ingestion and before any QC checks (rule-based, statistical, or AI-based). It ensures all downstream QC modules receive complete data and function correctly. In financial data pipelines, missingness is rarely completely at random (MCAR); it is often missing at random (MAR) due to observable factors, or missing not at random (MNAR) due to unobserved market behavior or reporting practices. Conventional techniques such as listwise deletion or mean imputation can lead to significant information loss or introduce systematic bias. To address these limitations, the proposed QC framework includes an imputation-aware, model-based QC layer designed to treat imputation as a governed process rather than a one-time correction.

The design follows three guiding principles:

1. **Missingness Profiling (Upstream QC):** Null rates and delta-null metrics are logged at regular intervals, and statistical tests are applied to classify missingness patterns as MCAR, MAR, or MNAR. This profiling step prevents uncontrolled propagation of data gaps to downstream stages.

2. **Method Portfolio (Model-Based QC):** Depending on the data type and identified missingness pattern, the framework selects from a portfolio of imputation methods. These include MissForest(Stekhoven & Bühlmann, 2012) for mixed-type tabular data, MICE(Van Buuren & Groothuis-Oudshoorn, 2011) for approximately linear relationships, SoftImpute(Hastie et al., 2015) for low-rank financial panels, and temporal deep learning models such as GRU-D for irregular time series. Random Forest–based imputers are particularly advantageous because they can accommodate missingness while supporting concurrent anomaly detection (Harvey et al., 2025).

3. **Residual and Uncertainty Checks (Downstream QC):** After imputation, lightweight predictive QC models estimate expected values and corresponding confidence intervals. Records are flagged when residuals exceed acceptable limits and the associated uncertainty is low. This predict-then-check mechanism transforms imputation from a black-box correction into a transparent and auditable safeguard.

This integrated design elevates imputation to a QC intervention, improving stability in drift detection, reducing false alarms, and strengthening trust in downstream analytical and modeling processes. The subsequent section presents a case study focusing on the rule-based and AI-based QC components discussed in Sections 3.2.1 and 3.3.

### 3.5 QC Notifications and Visualizations

In real-world applications, Data QC would be created by model owners often in partnership with engineers to verify the accuracy of production analytics and all associated data sources regularly with predetermined (daily, weekly, monthly, etc.) frequency. This process may be automated and can be managed by a dedicated team to ensure consistency and scalability. In an automated QC framework, following the execution of each data or model job, a corresponding QC job is triggered. If any threshold is breached during the quality control process, a notification is sent in real time and breach values are saved in a specific folder for review and resolution of data breaches. The subsequent job can only start once the previous quality control job has either run successfully or been manually signed off as a false alarm. The next job will run automatically, in the absence of breaches.

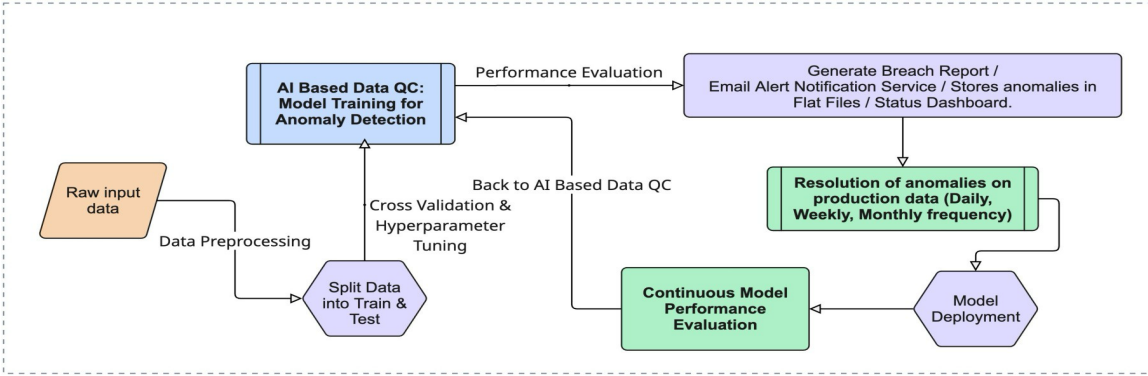

*Figure 3.* AI Based Data QC Flow Chart

### 3.6 The Complete System Integration

The unified QC system operates as an orchestrated pipeline that connects the core Python based library to a front-end code and model-level components through Airflow/internal cronjob workflows. Each task—data ingestion, validation, and QC is containerized and runs on a hybrid cloud/linux based on-prem environment. Configurations are fully YAML/JSON-driven, allowing environment-specific customization without redeployment. The orchestration supports automatic retries, job restarts, and check-pointing for resilience. The architecture scales horizontally across N data sources, enabling parallel execution with linear performance gains and centralized monitoring via internal dashboarding tool similar to Grafana.

## 4 CASE STUDY

We present a case study on the Rule and Statistical Based QC checks performed on the "Corporate Bonds Indices" dataset on Kaggle (DenzilG, 2023). We used this dataset to demonstrate the proposed framework for rule-based and statistical QC checks. The dataset comprises financial time series data from 1998 onward, covering corporate bond indices across multiple sectors and regions. It serves as a representative benchmark for assessing bond performance, risk exposure, and market dynamics. The dataset covers multiple indices, 21 numerical and 1 categorical features, making it longitudinal and multi-period, similar to many financial monitoring systems making it a good representative of real-world finance datasets for data-quality control and anomaly detection case study.

Additionally, we demonstrated the supervised and unsupervised AI methods to detect outliers using one of the public ADBench benchmark datasets (S.Han et al., 2022), Fraud Dataset (Jiang et al., 2022). The ADBench benchmark consists of 57 tabular datasets drawn from diverse application domains (finance, healthcare, image, NLP), providing for each dataset the number of samples, features, anomaly ratio, and enabling evaluation of supervised, semi-supervised and unsupervised anomaly detection methods. More specifically, the Fraud dataset contains about 284807 samples with 29 features and only 0.17% (492) anomalies; making it a heavily imbalanced dataset.

### 4.1 Rule-Based Validation and Statistical QC

Several standard QC checks were implemented within the framework: `do_null_level` to compute and report missing value counts; `do_outlier_detection_std` to flag outliers exceeding mean-based thresholds derived from standard deviation; `do_outlier_detection_range` to detect values lying outside historical minimum and maximum bounds computed over ten prior observations per feature; `do_positive_only` to ensure all numerical entries are non-negative; and `do_last_value_delta_check` to identify features that remain constant across consecutive periods.

Among these, the framework detected a breach only in the `do_outlier_detection_range` check, indicating potential anomalies in bond index values that deviated from recent historical behavior. The outcomes of all QC checks, along with their functional status, are recorded in a "Status File", as shown in Figure 8 in the appendix. This file logs key metadata for each run, including the execution date, check name, timestamp, and overall success indicator of the QC process. When a breach is detected, a detailed report is automatically generated and shared with relevant stakeholders. The report includes the type of breach, its file-system location, the corresponding input data source, and the specific column and values affected. The `BREAK_-THE_PROCESS` flag governs the workflow response. When set to `false`, the process continues execution and records a yellow status, signaling a non-critical issue. When set to `true`, the pipeline halts and records a red status until the data team investigates, resolves the breach, and resumes processing. Please refer to Figure 9 in the appendix for email format.

### 4.2 AI Based Advanced QC

Fundamentally, the PyOD library has been integrated into the QC framework to perform outlier detection on downstream data with benchmarking done using ADBench datasets to verify the performance of the anomaly algorithms used. For illustration purposes, we randomly sampled an equal number of instances (100) from each class in the dataset, shuffled the data and extracted the resulting features and labels. The Extreme Boosting Outlier Detection (XGBOD) (Zhao & Hryniewicki, 2018) model was then used to perform anomaly detection. We selected XGBOD because it combines the strengths of both supervised and unsupervised machine learning methods creating a hybrid approach which uses each of their capabilities to perform outlier detection.

A simple grid search is performed using GridSearchCV (Pedregosa et al., 2011) to tune hyperparameters. The training data was split into 80% train and 20% test, and a threshold was set at 0.2 to convert the anomaly scores into binary labels (1 for anomalies, 0 for non-anomalies). We used the same dataset for anomaly detection using Isolation Forest which identifies instances that deviate from the norm through isolation. The model was trained on unlabeled data, by skipping the existing labels in the original data. Figure 4 and 5 show the results by comparing the predicted anomalies returned from the XGBOD and Isolation Forest model. This demonstrates that the integrated AI-based QC aspect of the framework is effective in identifying anomalies even in highly imbalanced datasets.

|           | 0        | 1        | accuracy | macro avg | weighted avg |
|-----------|----------|----------|----------|-----------|--------------|
| precision | 0.888889 | 0.909091 | 0.9      | 0.89899   | 0.9          |
| recall    | 0.888889 | 0.909091 | 0.9      | 0.89899   | 0.9          |
| f1-score  | 0.888889 | 0.909091 | 0.9      | 0.89899   | 0.9          |
| support   | 9.000000 | 11.000000| 0.9      | 20.00000  | 20.0         |

*Figure 4.* Supervised AI Results Using XGBOD for Anomaly Detection on Fraud Dataset.

The recall for the fraudulent class was high, meaning the model detected the most fraudulent cases in the test set. However, the precision was lower, indicating that many normal transactions were misclassified as fraud. This is typical in imbalanced datasets, where maximizing recall often comes at the cost of precision. For illustration, we did not conduct extensive threshold tuning or employ advanced techniques like ROC curve analysis or Precision-Recall curve optimization, which could improve the model's precision and recall balance.

|           | 0         | 1         | accuracy | macro avg | weighted avg |
|-----------|-----------|-----------|----------|-----------|--------------|
| precision | 0.482759  | 1.000000  | 0.5      | 0.741379  | 0.758621     |
| recall    | 1.000000  | 0.062500  | 0.5      | 0.531250  | 0.500000     |
| f1-score  | 0.651163  | 0.117647  | 0.5      | 0.384405  | 0.366621     |
| support   | 14.000000 | 16.000000 | 0.5      | 30.000000 | 30.000000    |

*Figure 5.* Unsupervised AI Results Using Isolation Forest for Anomaly Detection on Fraud Dataset

To quantify the quality gains achieved through missing-value imputation, we evaluated *unsupervised* AI-based QC pipelines on the ADBench Fraud dataset (Jiang et al., 2022). The objective was to measure how imputation-aware feature completion influences detection quality and false positive reduction in automated QC checks. The model was evaluated using standard metrics: precision, recall, F1-score, and false-positive rate (FPR).

*Table 1.* Comparison of QC performance before and after imputation-aware training.

| Model Type                        | Precision | Recall | F1-Score | Approx. FPR |
|-----------------------------------|-----------|--------|----------|-------------|
| Unsupervised QC (baseline)        | 0.74      | 0.53   | 0.38     | ∼0.48       |
| Unsupervised QC (with imputation) | 0.90      | 0.90   | 0.90     | ∼0.10       |

The baseline unsupervised detector (trained directly on incomplete data) achieved moderate precision (0.74) and recall (0.53), indicating high sensitivity to noise and missing-value distortions. The corresponding false-positive rate was nearly 48%, reflecting frequent mis-classification of incomplete but valid records as anomalies.

Introducing imputation-aware pre-processing significantly improved both precision and recall to approximately 0.9, while reducing the false-positive rate to below 10%. This corresponds to a ∼5× reduction in false alerts and a relative F1-score improvement of more than 130%. The improvement arises primarily from imputing likely values for sparse or missing attributes before anomaly scoring.

Operationally, this translates to far fewer false QC triggers in production, less manual review effort, and faster data release cycles. This demonstrates that integrating imputation awareness into AI-based QC pipelines not only enhances accuracy but also meaningfully reduces alert fatigue, thereby increasing trust in automated data validation workflows.

### 4.3 System Performance

We evaluate system performance across key dimensions of scalability, efficiency, and resilience on an internal dataset representative of real production workloads where the QC system is currently deployed. Specifically, we measure (i) end-to-end processing latency, (ii) throughput speedup over sequential baselines, (iii) scalability with increasing number of data sources, (iv) fault recovery time, and (v) resource utilization during steady-state operation.

*Table 2.* Summary of performance metrics for the production system.

| Metric | Value / Result | Observation |
| --- | --- | --- |
| End-to-end latency (10 GB input) | 8.6 min | Low-latency processing |
| Throughput speedup | 3× vs. sequential | High efficiency |
| Scalability (100 sources) | 84% parallel efficiency | Near-linear scaling |
| Fault recovery time | < 30 s | Rapid restart |
| CPU utilization | ≈60% | Stable usage |
| Orchestration overhead | < 8% | Minimal cost |

The system demonstrates production-grade scalability, efficiency, and fault tolerance. Distributed execution scales nearly linearly up to 100 data sources, maintaining 84% parallel efficiency. Automated fault recovery restarts failed jobs within 30 seconds, ensuring high availability, while CPU utilization remains stable at 60% with less than 8% orchestration overhead.

## 5 CONCLUSION

We presented a unified AI system for Data QC and DataOps management designed for regulated environments, with finance as a representative domain. The system integrates rule-based validation, statistical QC, AI-based anomaly detection, and imputation-aware QC into a continuously governed layer that operates across ingestion, model pipelines, and downstream consumption.

Our implementation demonstrates that treating data quality as a first-class system concern yields practical benefits in production workflows. The framework operationalizes QC-driven observability, reduces manual triage through automated breach handling and notifications, and improves auditability with immutable run metadata, lineage, and policy versioning. Case studies on publicly available datasets of time-series bond indices and transactional fraud data show that rule-based and statistical checks catch schema and range violations early, while AI-based components identify context-dependent anomalies and support intelligent imputation.

The system is engineered for real deployments. It composes established open-source tools with custom modules, supports cloud storage and schedulers, and exposes configuration-driven policies so thresholds and actions can be updated without redeployments. We provide reproducible artifacts, including configuration templates, orchestrated pipelines, and scripted evaluations for both tabular and streaming tabular data, to facilitate independent validation and adoption.

Though we have not yet present large-scale distributed benchmarks across heterogeneous clusters, formal guarantees (e.g., formal verification and certifications) on false alert rates under concept drift, or results on unstructured modalities such as text and documents common in compliance workflows, these are natural extensions for future work. Future work will focus on three directions: scaling out distributed QC execution for high-throughput feeds, extending the framework to multimodal data with text and document validators integrated into the same governance plane, and adding adaptive policies that learn QC thresholds online under changing market regimes. We also plan to release a lightweight benchmark and a reproducibility checklist tailored to data quality systems to help standardize evaluation in this space.

In summary, unifying rule-based, statistical, AI-based, and imputation-aware QC within a governed DataOps layer provides a practical path to trustworthy, auditable, and resilient data pipelines in regulated settings. We expect this systems perspective to be immediately useful to practitioners and to serve as a foundation for future work on scalable, learning-based data quality management.

## ACKNOWLEDGEMENTS

The views expressed in this work are those of the authors alone and not of BlackRock, Inc.

# APPENDIX

*Table 3.* Common Types of Data Quality Issues in the Financial Sector

| Type of Issue | Example | Impact on Decision-making |
|---|---|---|
| Incomplete Data | Missing customer income details in loan applications | Skewed risk assessment; reduced model accuracy |
| Inaccurate Data | Incorrect transaction amounts due to entry errors | Misleading financial insights; flawed forecasting |
| Inconsistent Data | Different date formats across databases (MM/DD/YYYY vs. DD/MM/YYYY) | Data integration challenges; potential misclassification |
| Duplicate Data | Multiple records of the same client in CRM systems | Inflated customer base; redundancy in analysis |
| Outliers/Anomalies | Unusually large transaction amounts not representative of customer behavior | Model overfitting; false positives in fraud detection |
| Corrupt/Noisy Data | Data loss during transfer or corrupted files | Unusable datasets; compromised model training |
| Timeliness Issues | Delayed reporting of stock prices or credit scores | Outdated decisions; missed risk signals |
| Relevance Issues | Inclusion of outdated loan history not applicable to current analysis | Ineffective predictions; unnecessary computational costs |
| Biased Data | Historical data reflecting discriminatory lending practices | Reinforcement of unfair bias; regulatory compliance risks |

*Table 4.* Overview of aggregated and individual columnar QC checks

| Check Name | Data QC Rules | Examples |
|---|---|---|
| Aggregate-count | File count should be within a certain specified threshold | Count of rows present in the file should be between 60,000 and 80,000 |
| Aggregate-null-diff-percentage-change | Calculate the null value percentage difference between yesterday's and today's data in a particular column | Percentage change between today and yesterday null values of "volume" column should be below 10% |
| Aggregate-mean-diff-percentage-change | Calculate the mean value percentage difference between yesterday and today's data | Percentage change between today and yesterday's mean values of "volume" column should be below 10% |
| Aggregate-correlation-change | Correlation for a given column between yesterday and today's data should be within a specified threshold | Correlation between today and yesterday's "volume" column should be above 0.9 |
| Aggregate-col-null-values | Percentage of null values present in a column should be within a specified threshold | Percentage of null values in "volume" column should be below 15% |
| Aggregate-common-cusips-change | Percentage of common cusips that have changed values day over day for a given column | Percentage change in the value of "rating" column over common cusips between today's and yesterday's data should be below 5% |
| Aggregate-compare-col-names | Column names should remain the same across days | Column names should be constant comparing today's and yesterday's data |
| Security-null-value | Check for null values in a column which should not be present | "Date" column should not have any null values |
| Security-dup-value | Duplicate values should not be present in a column | "cusip" column should only contain unique values |
| Security-col-value | Values present in a column should be within a specified threshold | Each "cusip" amount issued should be between certain values |
| Security-col-ratio-check | Ratio of two columns should be less than the specified threshold | Ratio of "volume" by "amount issue" column should be less than 10% |

| Security-multiday-check | Check if the value of a column for two consecutive days is constant | Sum of a column at "cusip" level should not be 0 |
| --- | --- | --- |
| Security-missing-file-check | Check that the required file is present and has required data | File for the current date should not be empty or missing, otherwise send an alert |

*Table 5.* Overview of aggregated and individual columnar QC checks

| Supervised | Extreme Boosting Based Outlier Detection (XG-BOD) (Zhao & Hryniewicki, 2018) | Using labelled data points, distinguish between normal data points and anomalous points with precision |
| --- | --- | --- |
| Supervised | Random Forest Model | Using historically labelled data points, the model identifies exceptions as either true positives or false positives on new datasets |
| Supervised | Multiple Linear Regression Analysis | Forecast returns of securities based on different factors by calculating volatility relative to overall market price |
| Unsupervised | Isolation Forests | Outlier detection on unlabelled datasets by efficiently isolating anomalies in high-dimensional data |
| Unsupervised | Kullback-Leibler Divergence (KL Divergence) (Kurian & Allali, 2024) | Capture drift and skewness on data points to assess divergence between incoming and training data distributions |
| Unsupervised | Principal Component Analysis (PCA) (Jolliffe, 2011) | Used jointly with clustering methods for transaction monitoring in fraud detection |
| Semi-supervised | Generative Adversarial Networks (GANs) (Ian et al., 2014) | Detect anomalous data points by identifying whether a datapoint is real or synthetically generated |
| Reinforcement Learning | Gated Deep Q-networks (Hasselt et al., 2016) | Automatically detect and resolve data inconsistencies on large financial datasets |

### Centralized QC

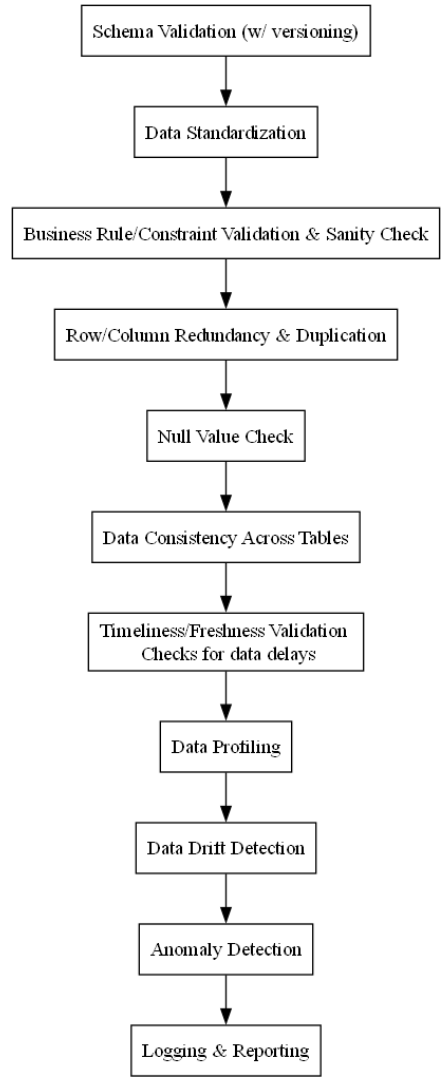

*Figure 6.* A comprehensive Centralized Data QC system

### Model-Level QC

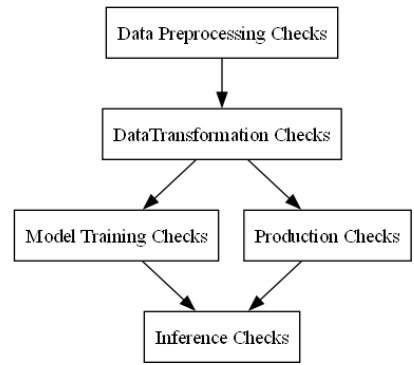

*Figure 7.* Steps in a model-Level QC

## QC Breach Status File

| Series Run Date | Check | Status Update Timestamp | Status |
|---|---|---|---|
| 20111220 | Missing Value Check | 27/09/2024 15:03 | Success - No Breach Detected |
| 20111220 | Positive Values Only | 27/09/2024 15:03 | Success - No Breach Detected |
| 20111220 | Outlier Check - Std-Dev Range | 27/09/2024 15:03 | Success - No Breach Detected |
| 20111220 | Outlier Check - Min-Max Range | 27/09/2024 15:03 | Success - Breach Detected |
| 20111220 | Value Delta Change Check | 27/09/2024 15:04 | Success - No Breach Detected |
| 20111227 | Missing Value Check | 27/09/2024 14:01 | Success - No Breach Detected |
| 20111227 | Positive Values Only | 27/09/2024 14:01 | Success - No Breach Detected |
| 20111227 | Outlier Check - Std-Dev Range | 27/09/2024 14:01 | Success - No Breach Detected |
| 20111227 | Outlier Check - Min-Max Range | 27/09/2024 14:01 | Success - Breach Detected |
| 20111227 | Value Delta Change Check | 27/09/2024 14:01 | Success - No Breach Detected |
| 20111228 | Missing Value Check | 27/09/2024 13:57 | Success - No Breach Detected |
| 20111228 | Positive Values Only | 27/09/2024 13:57 | Success - No Breach Detected |
| 20111228 | Outlier Check - Std-Dev Range | 27/09/2024 13:57 | Success - No Breach Detected |
| 20111228 | Outlier Check - Min-Max Range | 27/09/2024 13:57 | Success - Breach Detected |
| 20111228 | Value Delta Change Check | 27/09/2024 13:57 | Success - No Breach Detected |

*Figure 8.* Status File representing the check run date, name, status-update timestamp and final status

## QC Breach Status Email

**QC Failure Report** — **NOT BREAK THE PROCESS CHECKS**

**Aggregate Count Change Check**
- QC Breach Path:
- Current Path:
- Previous Path:
- Assertion Query: Day-Over-Day non-null row count change is more than 10.0
- Assertion Description: Non-null row count change between yesterday and today's data
- Sample Invalid: **Today's Date Change in Row Count**
  2025-08-22   3154

**Correlation Change Check**
- QC Breach Path:
- Current Path:
- Previous Path:
- Assertion Query: Correlation Threshold is defined at 0.8
- Assertion Description: Correlation for the given portfolios between yesterday and today's data
- Sample Invalid: **Today's Date Correlation**
  2025-08-22   0.014354

**Value Difference Check**
- QC Breach Path:
- Current Path:
- Previous Path:
- Assertion Query: Value should not differ more or less than 30.0 percent
- Assertion Description: Value change for the 2 portfolios between yesterday and today

*Figure 9.* Status Breach Report highlighting the severity of the breaches. The frequency of the report can be customized to be set hourly, daily, weekly or monthly